\documentclass[a4paper,10pt]{article}

\usepackage{amsmath}
\usepackage{amssymb}
\usepackage{amsthm}
\usepackage{amsfonts}
\usepackage{xcolor}
\usepackage{graphicx}
\usepackage{mathtools}
\usepackage{enumerate}
\usepackage{multirow}
\usepackage{rotating}
\usepackage{setspace}
\usepackage{lineno}
\usepackage{url}
\usepackage{subfigure}

\theoremstyle{theorem}

\theoremstyle{definition}

\newcommand\num{\addtocounter{equation}{1}\tag{\theequation}}

\usepackage{geometry}

\begin{document}
\title{ Control Intervention Strategies for Within-Host, \\ Between-Host and their Efficacy in the Treatment, Spread of COVID-19 :  A Multi Scale Modeling Approach }
	
\vspace{0.1in}
\author{ Bhanu Prakash D$^{a}$,  D. K. K. Vamsi$^{a*}$ ,  D. Bangaru Rajesh$^{a}$, Carani B Sanjeevi$^{b, c}$  \\\\ \medskip $^{a}$Department of Mathematics and Computer Science, Sri Sathya Sai Institute of Higher Learning - \\ SSSIHL, India  \\ $^{b}$ Vice-Chancellor, Sri Sathya Sai Institute of Higher Learning -  SSSIHL, India  \\  $^{c}$ Department of Medicine, Karolinska Institute, Stockholm, Sweden \\ $^{*}$ \textit{Corresponding Author} \\\\ \medskip bhanuprakash@sssihl.edu.in,    dkkvamsi@sssihl.edu.in, bangaru.dmacs@gmail.com, \\ sanjeevi.carani@sssihl.edu.in, sanjeevi.carani@ki.se}
	\date{}
	\maketitle
	
	\begin{abstract} \vspace{.25cm}
	
	The COVID-19 pandemic has resulted in more than 14.5 million infections and 6,04,917 deaths in 212 countries over the last few months. Different drug intervention acting at multiple stages of pathogenesis of COVID-19 can substantially reduce the infection induced,thereby decreasing the mortality. Also population level control strategies can reduce the spread of the COVID-19 substantially. Motivated by these observations, in this work we propose and study a  multi scale model linking both within-host and between-host dynamics of COVID-19. Initially  the natural history dealing with the disease dynamics is studied. Later, comparative effectiveness is performed to understand the efficacy of  both the within-host and population level interventions. Findings of this study suggest that a combined strategy involving  treatment with drugs such as Arbidol, remdesivir, Lopinavir/Ritonavir that inhibits viral replication  and immunotherapies like monoclonal antibodies, along with environmental hygiene and generalized social distancing  proved to be the best and optimal in reducing the basic reproduction number and environmental spread of the virus at the population level.  \vspace{.35cm}
	
	\end{abstract}
	

	\vspace{.12 in}
	
	The pandemic COVID-19, caused by SARS-CoV-2, spread its tentacles through out the world by taking lives of 6,04,917 people and livelihood of many more across the globe. Research communities across the world are racing against time in contributing their piece of knowledge in tackling this virus \cite{link1}. 

Mathematical models play a crucial role in this journey as they are helpful in multi-fold. Firstly, they help in Understanding the dynamics of infection and its spread in the society. They help in  studying the success of various control measures that can be implemented in order to avoid further damage. Secondly, the within-host mathematical modeling helps to study the  dynamics of virus in the human body and can help us in understanding efficacy of different drug intervention acting at multiple stages of pathogenesis which in turn  can help in identification of potential vaccine candidates. Some of the works dealing with population level studies and within-host studies  for diseases such as Dengue, HIV, Influenza include  \cite{multi1,p2,hiv1} and references within. Recent works dealing with  population level studies and within-host studies for COVID-19 can be found in \cite{pop1,p3,pop2,mathbio,inhost2}

In addition to these, there are multi scale models linking within-host and between-host population scales which helps us understand not only the efficacy of the intervention at individual level but also the effectiveness at the population level. Broadly  there are five different categories of multi scale models. As enlisted in \cite{multi1}, they are Individual-based multi scale model, nested multi scale model, embedded multi scale model, hybrid multi scale model and coupled multi scale model. Few works involving multi scale modeling approaches for diseases  include \cite{multi1,multi2}. To the best of our knowledge, there is no work dealing with multi scale modeling approach for Covid-19.

 Motivated by the above in this study, we propose and study a nested multi scale model. Using the technique of  comparative effectiveness we study the efficacy of drug interventions at individual level and population level control measures. The multi scale model and the corresponding  interventions study  which is being attempted here is the first of its kind for Covid-19.

\section{The Multi Scale Model Formulation} \vspace{.25cm}

The  multi scale model for COVID-19 disease dynamics across two scales that are  within-host  and between-host. The  model consists of eight compartments involving susceptible epithelial cells $S_{h}$, infected epithelial cells $I_{h}$, SARS-CoV-2 viral load $V_{h}$ at within-host scale and susceptible human population $S_{p}$, exposed population $E_{p}$, infected population $I_{p}$, recovered populations $R_{p}$ and environmental viral load $V_{p}$ at between-host scale.  The  assumptions for the proposed model include the following.

1. The within-host dynamics are assumed to occur at fast time scale $s$ while the dynamics of the between-host scale variables is assumed to occur at slow time scale $t.$

2. We assume that SARS-CoV-2 virus can be transmitted only through environmental(indirect) transmission. 

3. We do not consider the asymptomatic patients for this study as there is no sufficient evidence of within-host dynamics of asymptomatic patients. We  employ the basic SEIR model at between-host level.

Based on the above assumptions we propose the  following multi scale model for COVID-19.
\begin{align*}
\frac{\mathrm{d} S_{h}(s)}{\mathrm{d} s} &=  -\beta S_{h}(s)V_{h}(s) \\
\frac{\mathrm{d} I_{h}(s)}{\mathrm{d} s} &=  \beta S_{h}(s)V_{h}(s)-(d_{1}+d_{2}+d_{3}+d_{4}+d_{5}+d_{6}) I_{h}(s)-\mu I_{h}(s) \\
\frac{\mathrm{d} V_{h}(s)}{\mathrm{d} s} &=  \alpha I_{h}(s)-(b_{1}+b_{2}+b_{3}+b_{4}+b_{5}+b_{6})V_{h}(s)-\alpha _{h}V_{h}(s) \\
\frac{\mathrm{d} S_{p}(t)}{\mathrm{d} t} &=  \pi _{p} - \mu _{p}S_{p}(t) -\frac{\eta _{p} S_{p}(t)I_{p}(t)}{N_{p}(t)}-\eta _{w}S_{p}(t)V_{p}(t) \\
\frac{\mathrm{d} E_{p}(t)}{\mathrm{d} t} &=  \frac{\eta _{p} S_{p}(t)I_{p}(t)}{N_{p}(t)}+\eta _{w}S_{p}(t)V_{p}(t)-(\omega _{p}+\mu _{p})E_{p}(t) \\
\frac{\mathrm{d} I_{p}(t)}{\mathrm{d} t} &=  \omega _{p}E_{p}(t) -(\tau _{p}+\mu _{p})I_{p}(t) \\
\frac{\mathrm{d} R_{p}(t)}{\mathrm{d} t} &=  \tau _{p}I_{p}(t)-\mu _{p}R_{p}(t) \\
\frac{\mathrm{d} V_{p}(t)}{\mathrm{d} t} &=  V_{h}(s)\alpha _{h}I_{p}(t) - \pi V_{p}(t) \num \label{eqn1} 
\end{align*}

The first compartment in the model \ref{eqn1} deals with the dynamics of susceptible epithelial cells S$_{h}$(s). They decrease at a rate $\beta$ following contact with the virus. The second compartment deals with the dynamics of infected epithelial cells. They are increased through infection of susceptible  cells and are decreased through clearance by cytokines and chemokines such as IL-6, TNF-$\alpha$, CCL5, CXCL8, CXCL10 at the rate $d_{1}$, $d_{2}$, $d_{3}$, $d_{4}$, $d_{5}$ and $d_{6}$ respectively and through natural death at a rate $\mu$. Compartment three deals with SARS-CoV-2 viral load. The viral load is increased due to reproduction of virus in the infected cells at the rate $\alpha$. This viral load is decreased due to clearance by cytokines and chemokines such as IL-6, TNF-$\alpha$, CCL5, CXCL8, CXCL10 at the rate $b_{1}$, $b_{2}$, $b_{3}$, $b_{4}$, $b_{5}$ and $b_{6}$ respectively and they are released into environment at a rate $\alpha_{h}$. The term $\alpha_{h}$ links within-host and between-host scales in a uni-directional way.

We assume that the total human population $N_{p}$ in the between-host dynamics is divided into four subgroups denoted by $S_{p}$, $E_{p}$, $I_{p}$ and $R_{p}$ which represent respectively, the susceptible, exposed, infected and recovered or the removed population. Compartment 4 in the model (\ref{eqn1}) describes the dynamics of susceptible humans. They are assumed to be supplied at a constant rate $\pi _{p}$ through birth and are removed at the natural death rate $\mu_{p}$. They are also reduced by interaction with infected cells and virus in the environment through the terms $\frac{\eta _{p} S_{p}(t)I_{p}(t)}{N_{p}(t)}$ and $\eta _{w}$S$_{p}$(t)V$_{p}$(t) respectively. Compartment 5  describes the dynamics of exposed humans. They become exposed due to contacts with either infected people or virus particles through the terms $\frac{\eta _{p} S_{p}(t)I_{p}(t)}{N_{p}(t)}$ and $\eta _{w}$S$_{p}$(t)V$_{p}$(t) respectively. They are removed at the natural death rate $\mu_{p}$ and the rate at which they become infected through $\omega _{p}$.
Compartment 6 describes the dynamics of infected humans. Exposed humans become infected after the incubation period at the rate $\omega_{p}$. They are reduced through medication or natural death at the rate $\tau_{p}$ and $\mu_{p}$ respectively.
Compartment 7  describes the dynamics of recovered individuals. Infected humans are recovered at the rate $\tau_{p}$ and are removed due to natural death at the rate $\mu_{p}$.
Finally compartment 8 describes the dynamics of viral load in the environment.The viral load in the environment is contributed by infected cells at the rate  $\alpha_{h}V_{h}$ through coughing/sneezing. The virus particles cannot live in environment without invading into host cell for long time. The virus will get killed at the rate $\pi$. The values for these various parameters is given in {table \ref{parameters table}}.

\begin{table}[htp!]
\begin{center}
\begin{tabular}{ | c | c | c | c | c | }
\hline
 \textbf{Variable} & \textbf{Description} & \textbf{Value} & \textbf{Units} & \textbf{Source}\\ 
 \hline
 $\pi_{p}$ & Birth rate & 294.91 & d$^{-1}$ & \cite{p3} \\  
 \hline
 $\mu_{p}$ & Natural mortality rate & $\frac{1}{76.79 \times 365}$ & d$^{-1}$ & \cite{p3} \\  
 \hline
 $\eta_{p}$ & Contact rate & 0.05 & d$^{-1}$ & \cite{p3} \\  
 \hline
 $\eta_{w}$ & Disease transmission coefficient & 0.000001231 & d$^{-1}$ & \cite{p3} \\  
 \hline
 $\omega_{p}$ & Incubation Period & 0.00047876 & d$^{-1}$ & \cite{p3} \\  
 \hline
 $\tau_{p}$ & Removal or recovery of I$_{p}$ & 0.09871 & d$^{-1}$ & \cite{p3} \\  
 \hline
 $\pi$ & Removal rate of virus from environment & 0.01 & d$^{-1}$ & \cite{p3} \\  
 \hline
 $\beta$ & Rate at which healthy Pneumocytes are infected & 0.55 & d$^{-1}$score$^{-1}$ & \cite{mathbio} \\  
 \hline
 $\mu$ & Natural death rate of Type II Pneumocytes & 0.11 & d$^{-1}$ & \cite{mathbio} \\  
 \hline
 $\alpha$ & Burst rate of virus particles & 0.24 & d$^{-1}$ & \cite{mathbio} \\  
 \hline
 $\alpha_{h}$ & shedding rate of virus from infected human & 5.36 & d$^{-1}$ & \cite{mathbio} \\  
 \hline
  & Rate at which infected Pneumocytes & & \\ d$_{1}$,d$_{2}$,d$_{3}$,d$_{4}$,d$_{5}$,d$_{6}$ & are removed because of the release of & 0.01533 & cell$^{-1}$ d$^{-1}$ & \cite{multi1} \\ & cytokines IL-6,TNF-$\alpha$,CCL 5,& & \\ & CXCL 8,CXCL-10,INF-$\alpha $ respectively &   & \\  
 \hline
  & Rate at which viral particles & & \\ b$_{1}$,b$_{2}$,b$_{3}$,b$_{4}$,b$_{5}$,b$_{6}$ & are removed because of the release of & 250 & cell$^{-1}$ d$^{-1}$ & \cite{multi1} \\ & cytokines IL-6,TNF-$\alpha $,CCL 5, & & \\ & CXCL 8,CXCL-10,INF-$\alpha $ respectively &  & \\  
 \hline
\end{tabular}
\caption{Table describing the parameter values}
\label{parameters table}
\end{center}
\end{table}

\newpage

The initial values for the different between-host human population  are listed in the table \ref{initial table}.

\begin{table}[htp!]
\begin{center}
\begin{tabular}{ | c | c | c | c | }
\hline
 \textbf{Variable} & \textbf{Description} & \textbf{Initial Values} & \textbf{Source}\\ 
 \hline
 S$_{h}$(s) & Susceptible target cells & 4 $\times$  10$^{8}$ & \cite{p2}\\  
 \hline
 I$_{h}$(s) & Infected target cells    & 0 & Assumed \\  
 \hline
 V$_{h}$(s) & Viral load within infected cells & 3.0 & \cite{p2} \\  
 \hline
 S$_{p}$(t) & Susceptible Individuals & 8065518 & \cite{p3} \\  
 \hline
 E$_{p}$(t) & Exposed Individuals & 200000 & \cite{p3} \\  
 \hline
 I$_{p}$(t) & Infected Individuals & 282 & \cite{p3} \\  
 \hline
 R$_{p}$(t) & Recovered Individuals & 0 & \cite{p3} \\  
 \hline
 V$_{p}$(t) & Community Viral load & 50000 & \cite{p3} \\  
 \hline
\end{tabular}
\caption{Table describing the initial values for between-host compartments}
\label{initial table}
\end{center}
\end{table}



\section{The reduced-order multi scale model} \vspace{.25cm}

There are two difficulties in working with the proposed multi scale model. \vspace{.25cm}

1. Time scale mismatch: The within-host scale is in terms of a fast time scale $s,$ while the between-host scale is in terms of a slow time scale $t.$ \vspace{.25cm}

2. Transient $V_{h}(s):$  $V_{h}(s)$ remains non-zero only for a short period since the infection remains only for few days.

As discussed in \cite{multi1} these problems can be overcome by changing the measure of host-infectiousness from $V_{h}(s)$ to a new quantity $N_{h}$ (the area under the viral load curve). In similar lines to reduction of order done in \cite{multi1}, we get the reduced multi scale model as 
\begin{align*}
\frac{\mathrm{d} S_{p}(t)}{\mathrm{d} t} &=  \pi _{p} - \mu _{p}S_{p}(t) -\frac{\eta _{p} S_{p}(t)I_{p}(t)}{N_{p}(t)}-\eta _{w}S_{p}(t)V_{p}(t) \\
\frac{\mathrm{d} E_{p}(t)}{\mathrm{d} t} &=  \frac{\eta _{p} S_{p}(t)I_{p}(t)}{N_{p}(t)}+\eta _{w}S_{p}(t)V_{p}(t)-(\omega _{p}+\mu _{p})E_{p}(t) \\
\frac{\mathrm{d} I_{p}(t)}{\mathrm{d} t} &=  \omega _{p}E_{p}(t) -(\tau _{p}+\mu _{p})I_{p}(t) \\
\frac{\mathrm{d} R_{p}(t)}{\mathrm{d} t} &=  \tau _{p}I_{p}(t)-\mu _{p}R_{p}(t) \\
\frac{\mathrm{d} V_{p}(t)}{\mathrm{d} t} &=  N_{h}\alpha _{h}I_{p}(t) - \pi V_{p}(t) \num \label{eqn3}
\end{align*}

where      

\begin{displaymath}
    N_{h} = \frac{\mathfrak{R_{0}}}{\beta} \left[ 1 - e^{-\mathfrak{R_{0}}} - \mathfrak{R_{0}} e^{-2\mathfrak{R_{0}}} \right] 
\end{displaymath}
 
with the initial conditions $S_{p}(0)$ $\geq 0,$  $E_{p}(0)$ $\geq$ $0,$ $I_{p}(0)$ $\geq$ $0,$  $R_{p}(0)$ $\geq$ $0$ and $V_{p}(0)$ $\geq$ $0.$

In similar lines to the work \cite{multi1}, the basic reproductive number for the within-host scale sub model in model \ref{eqn1} is given by
\begin{equation}
    \mathfrak{R_{0}} = \frac{\alpha \beta S_{0}}{(b_{1}+b_{2}+b_{3}+b_{4}+b_{5}+b_{6}+\alpha_{h})(d_{1}+d_{2}+d_{3}+d_{4}+d_{5}+d_{6}+\mu)}
    \num \label{eqn2}
\end{equation}

Adding the compartments for the sub model \ref{eqn3}, we get the dynamics of total population governed by the following equation.

\begin{displaymath}
    \frac{\mathrm{d} N_{p}(t)}{\mathrm{d} t} = \pi _{p} - \mu _{p}N_{p}(t)
\end{displaymath}

The feasible region for the model (\ref{eqn3}) is given by 

\begin{displaymath}
        \Omega = \bigg\{ \left( S_{p}(t), E_{p}(t), I_{p}(t), R_{p}(t) \right) \in \mathbb{R}_{+}^{4} : N_{p}(t) \leq \frac{\pi_{p}}{\mu_{p}}, \ V_{p}(t) \in \mathbb{R}_{+} : V_{p}(t) \leq \frac{N_{h} \alpha_{h}}{\pi} \frac{\pi_{p}}{\mu_{p}} \bigg\}
\end{displaymath}
    
\section{Stability Analysis of the reduced multi scale model} \vspace{.25cm}

In this section, we do the stability analysis for  the reduced multi scale model  \ref{eqn3}.

\subsection{Disease free equilibrium and $R_{0}$} \vspace{.25cm}

The disease-free equilibrium for the model  \ref{eqn3} is given by

\begin{displaymath}
    E^{0} = \left( S^{0}_{p}, E^{0}_{p}, I^{0}_{p}, R^{0}_{p}, V^{0}_{p} \right) = \left( \frac{\pi _{p}}{\mu _{p}}, 0, 0, 0, 0 \right) 
\end{displaymath}

We now calculate the basic reproduction number of the multi scale model \ref{eqn3} using the next generation matrix approach \cite{r0}. 

The Jacobian  evaluated at the disease-free equilibrium, $E^{0}$, is given by

\begin{displaymath}
    J(E^{0}) = \begin{bmatrix}
    -(\omega_{p}+\mu_{p}) & \eta_{p}            & \frac{\eta_{w}\pi_{p}}{\mu_{p}}\\
    \omega_{p}            & -(\tau_{p}+\mu_{p}) & 0 \\
    0                     & N_{h}\alpha_{h}     & -\pi\\
    \end{bmatrix}
\end{displaymath}
The J(E$^{0}$) can be decomposed into two matrices $F$ and $V$ such that $J(E^{0})=F-V,$ where $F$ is the transmission and non-negative matrix describing the generation of secondary infections and $V$ is the transition and non-singular matrix, describing the changes in individual states such as removal by death, recovery and excretion of SARS-CoV-2 into the environment by infected human in the community. Since the environment acts as a reservoir of the infective pathogen, we have,

\begin{displaymath}
    F = \begin{bmatrix}
    0 & \eta_{p}        & \frac{\eta_{w}\pi_{p}}{\mu_{p}}\\
    0 & 0               & 0 \\
    0 & N_{h}\alpha_{h} & 0 \\
    \end{bmatrix}
    ,
    V = \begin{bmatrix}
    (\omega_{p}+\mu_{p}) & 0                  & 0   \\
    -\omega_{p}          & (\tau_{p}+\mu_{p}) & 0   \\
    0                    & 0                  & \pi \\
    \end{bmatrix}
\end{displaymath}

The basic reproductive number is given by the spectral radius (dominant eigenvalue) of the matrix $F V^{-1}.$ So in this case, we have the basic reproduction number of the  system \ref{eqn3} to be 

\begin{equation}
    \mathcal{R}_{0} = \frac{\eta_{p}\omega_{p} + \sqrt{(\eta_{p}\omega_{p})^2 + \frac{4\eta_{w}\pi_{p}N_{h}\alpha_{h}\omega_{p}(\omega_{p}+\mu_{p})(\tau_{p}+\mu_{p})}{\mu_{p}\pi}}}{2(\omega_{p}+\mu_{p})(\tau_{p}+\mu_{p})} \num \label{eqn8}
\end{equation}

\textbf{Theorem 1: } The DFE $E^{0}$ of the system \ref{eqn3} is locally asymptotically stable if $\mathcal{R}_{0}  < 1.$

\textbf{Proof: } The Jacobian matrix evaluated at E$^{0}$ is given by \\
\begin{displaymath}
    J(E^{0}) = \begin{bmatrix}
    -\mu_{p} & 0 & -\eta_{p} & 0 & -\frac{\eta_{w} \pi _{p}}{\mu _{p}} \\
    0 & -(\omega_{p}+\mu_{p}) & \eta_{p} & 0 & \frac{\eta_{w} \pi _{p}}{\mu _{p}}   \\
    0 & \omega_{p} & -(\tau_{p}+\mu_{p}) & 0 & 0  \\
    0 & 0 & \tau_{p} & -\mu_{p} & 0 \\
    0 & 0 & N_{h}\alpha_{h} & 0 & -\pi \\
    \end{bmatrix}
\end{displaymath}

From the above jacobian matrix, it can be seen that the two of the eigen values  are given by -$\mu_{p}$ (repeated twice).  The rest can be obtained through the characteristic equation given below.
\begin{displaymath}
    \lambda^3+a_1\lambda^2+a_2\lambda+a_3 = 0,
\end{displaymath}
where 
\begin{align*}
    a_1 &= \pi+\omega_p+\tau_p+2 \mu_p\\
    a_2 &= (\omega_p+\mu_p)(\tau_p+\mu_p) + \pi(\omega_p+\tau_p+2 \mu_p) - \omega_p\eta_p \\
    a_3 &= (\omega_p+\mu_p)(\tau_p+\mu_p)\pi - \omega_p(\pi\eta_p + \frac{\eta_w \pi_p N_h \alpha_h}{\mu_p})\\
\end{align*}

When $\mathcal{R}_{0}$ < 1, $a_2$, $a_3$ are positive. Since $a_1$ has all positive terms, all the coefficients of the characteristic equation are positive when $\mathcal{R}_{0}$ < 1. Further, by  Routh-Hurtwiz criterion for a third order polynomial, we get that all the eigen values of the characteristic polynomial to be negative as $a_i's > 0,$ for $i = 1, 2, 3.$\cite{routh}. Hence the  DFE $E^{0}$ is locally asymptotically stable when $\mathcal{R}_{0} < 1$ as all the eigen values of the jacobian matrix are negative.

\subsection{Endemic equilibrium} \vspace{.25cm}

Let $\overline{E} = (\overline{S_{p}},\overline{E_{p}},\overline{I_{p}},\overline{R_{p}},\overline{V_{p}})$ denote the endemic equilibrium point of the   system \ref{eqn3}. To get the endemic equilibrium, we set the left-hand side of the equations of the model system \ref{eqn3} equal to zero and determine the nontrivial solution of the resulting algebraic equations, which gives 

\begin{align*}
    \overline{S}_{p} &= \frac{1}{\mu_{p}} [\pi_{p} - \frac{(\omega_{p}+\mu_{p})(\tau_{p}+\mu_{p})}{\omega_{p}} \overline{I_{p}}] , \\
    \overline{E_{p}} &= \frac{\tau_{p}+\mu_{p}}{\omega_{p}} \overline{I_{p}} , \\
    \overline{I_{p}} &= \frac{\pi_{p}\omega_{p}}{(\omega_{p}+\mu_{p})(\tau_{p}+\mu_{p})} - \frac{\pi\pi_{p}\mu_{p}}{\eta_{p}\mu_{p}\pi + \eta_{w}N_{h}\alpha_{h}\pi_{p}}, \\
    \overline{R_{p}} &= \frac{\tau_{p}}{\mu_{p}} \overline{I_{p}}, \\
    \overline{V_{p}} &= \frac{N_{h}\alpha_{h}}{\pi} \overline{I_{p}}.
    \num \label{eqn9}
\end{align*}
It can be seen from the definition of $\overline{I_{p}}$ that the endemic equilibrium $\overline{E} $ exists only when $\mathcal{R}_{0} > 1.$

\textbf{Theorem 2: } The endemic equilibrium $\overline{E} $ of the system \ref{eqn3} is locally asymptotically stable if $\mathcal{R}_{0}  > 1.$

\textbf{Proof: } It can be seen from the above discussions that the endemic equilibrium $\overline{E} $ exists only when $\mathcal{R}_{0}  > 1.$ And also from theorem 1 the DFE $E^{0}$ is unstable in such a case. Moreover, as the system \ref{eqn3} admits only two equilibria from the uniqueness and boundedness of solutions for the system \ref{eqn3} we conclude that the endemic equilibrium $\overline{E} $ is asymptotically stable whenever it exists $ i.e., $ whenever $\mathcal{R}_{0}  > 1.$ \vspace{.30cm}

\section{Numerical simulations} \vspace{.30cm}

The multi scale  model \ref{eqn1} is  uni-directionally coupled such that  only the within-host scale sub model influences the between-host scale sub model without any reciprocal feedback. Owing to this,  in this  section, we study and numerically illustrate the influence of the  four within-host scale sub model parameters, namely 
 $\beta$,   $\alpha$,    $d (= d_1 + \dots + d_6) $        and 
$b (= b_1 + \dots + b_6)$        on the between-host scale sub model variables ( $S_p$, $E_p$, $I_p$, $R_p$, $V_p$).

\newpage
\begin{figure}
\begin{subfigure}
  \centering
  \includegraphics[width=7cm]{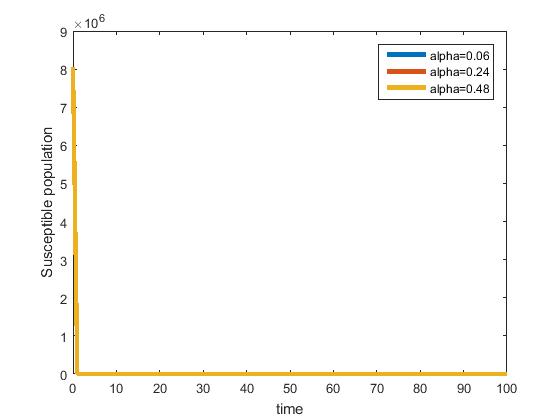}  
\end{subfigure}
\begin{subfigure}
  \centering
  \includegraphics[width=7cm]{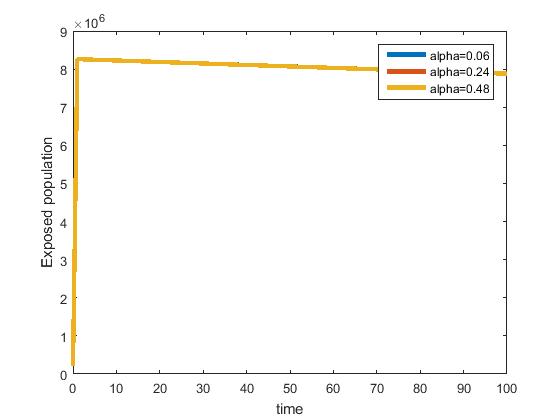}  
\end{subfigure}
\caption*{ }
\end{figure}

\begin{figure}
\begin{subfigure}
  \centering
  \includegraphics[width=7cm]{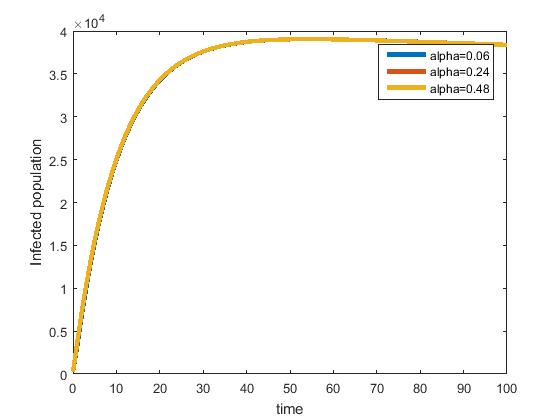}  
\end{subfigure}
\begin{subfigure}
  \centering
  \includegraphics[width=7cm]{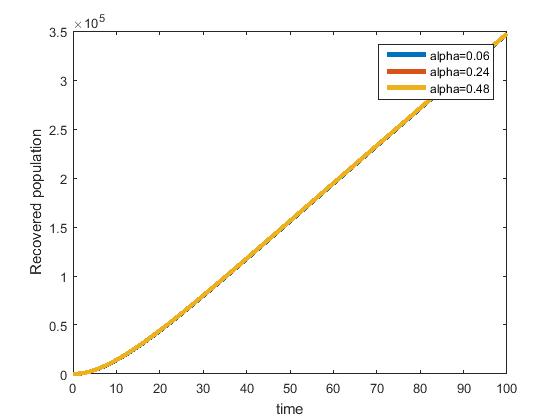}  
\end{subfigure}
\caption*{ }
\end{figure}

\begin{figure}[ht!]
\centering
\includegraphics[width=7cm]{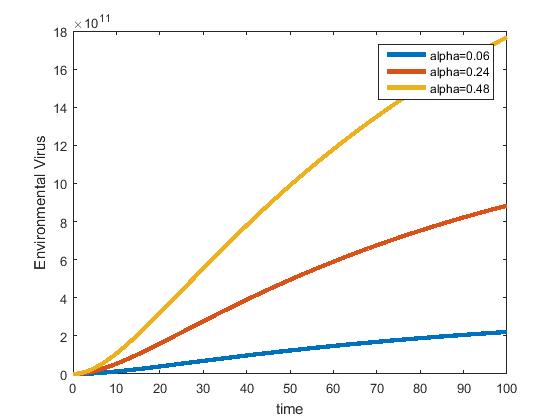}
\caption{Effects of variation of production rate of virus from within-host infected cells ($\alpha$).} \label{alpha}
\end{figure} \vspace{.35cm}

Figure \ref{alpha} depicts that as the  infected cell burst rate increases, SARS-CoV-2 transmission in the community also increases. Therefore,  drugs that inhibit viral replication (such as Arbidol, remdesivir, Lopinavir/Ritonavir) which in turn reduce the  production rate  of virus at within-host scale will likely reduce transmission of SARS-CoV-2 at between-host scale.

\newpage

\begin{figure}
\begin{subfigure}
  \centering
  \includegraphics[width=7cm]{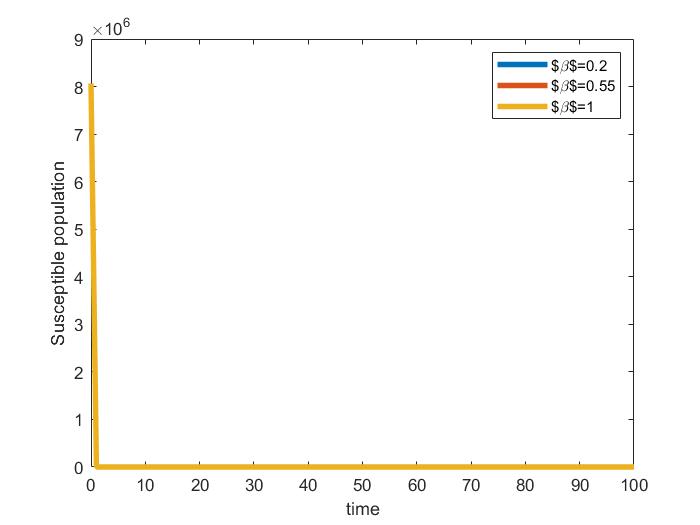}  
\end{subfigure}
\begin{subfigure}
  \centering
  \includegraphics[width=7cm]{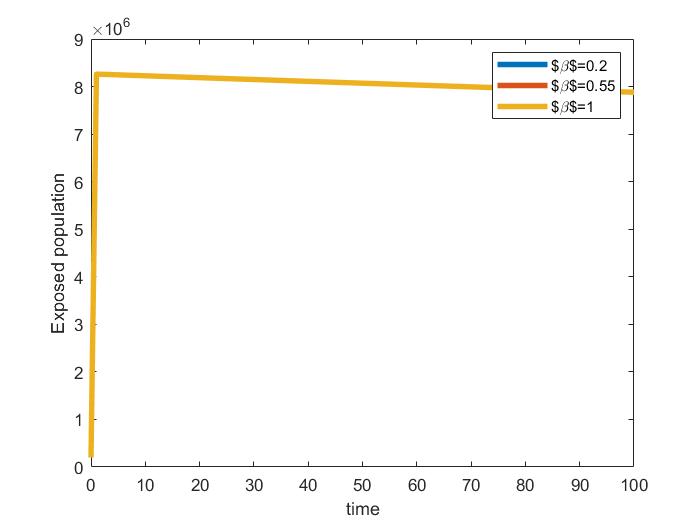}  
\end{subfigure}
\caption*{ }
\end{figure}

\begin{figure}
\begin{subfigure}
  \centering
  \includegraphics[width=7cm]{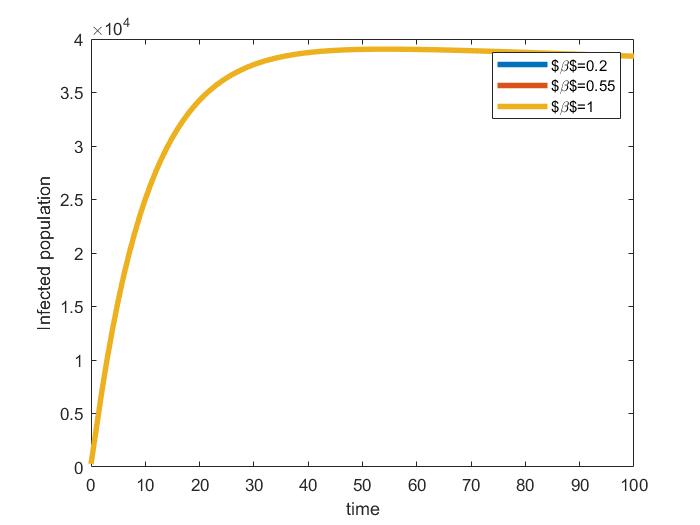}  
\end{subfigure}
\begin{subfigure}
  \centering
  \includegraphics[width=7cm]{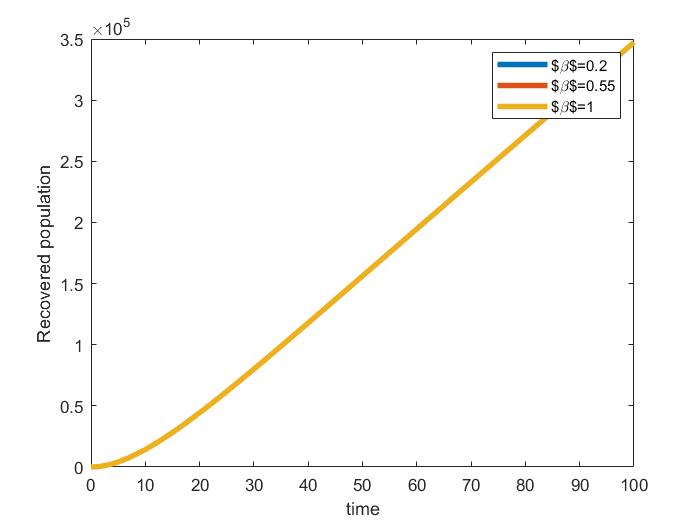}  
\end{subfigure}
\caption*{ }
\end{figure}

\begin{figure}[ht!]
\centering
\includegraphics[width=7cm]{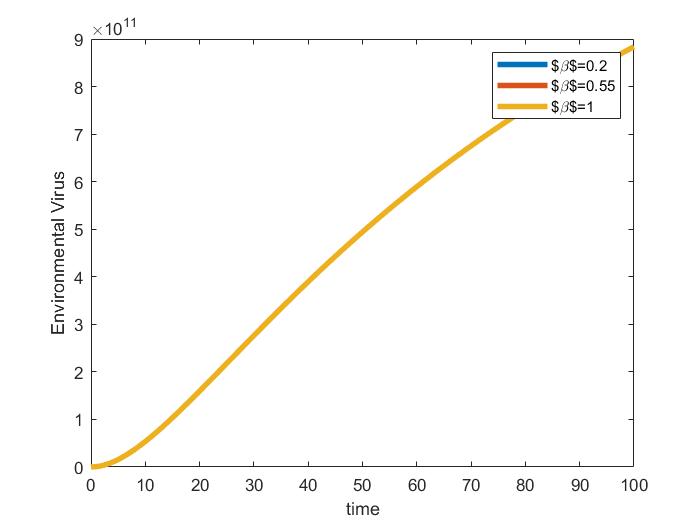}
\caption{Effects of variation of infection transmission probability ($\beta$).}\label{beta}
\end{figure} \vspace{.35cm}

Figure \ref{beta} depicts that the antiviral drugs (such as Hydroxychloroquine (HCQ))  that reduce infection rate of susceptible epithelial cells may have individual level benefits but have insignificant population level benefits.

\newpage

\begin{figure}
\begin{subfigure}
  \centering
  \includegraphics[width=7cm]{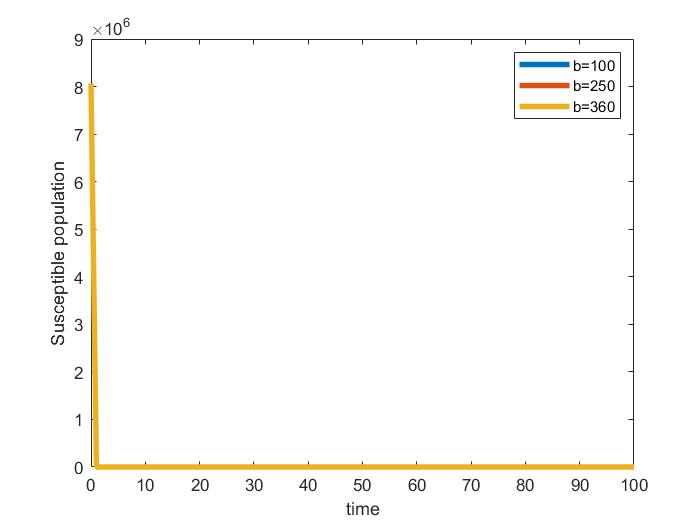}  
\end{subfigure}
\begin{subfigure}
  \centering
  \includegraphics[width=7cm]{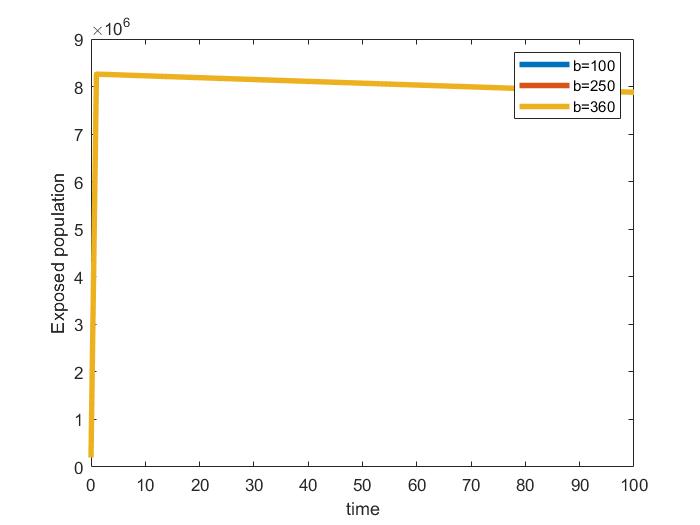}  
\end{subfigure}
\caption*{ }
\end{figure}

\begin{figure}
\begin{subfigure}
  \centering
  \includegraphics[width=7cm]{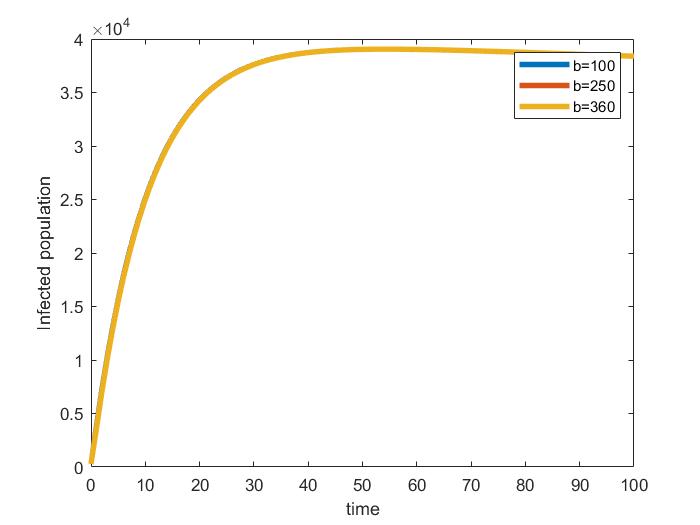}  
\end{subfigure}
\begin{subfigure}
  \centering
  \includegraphics[width=7cm]{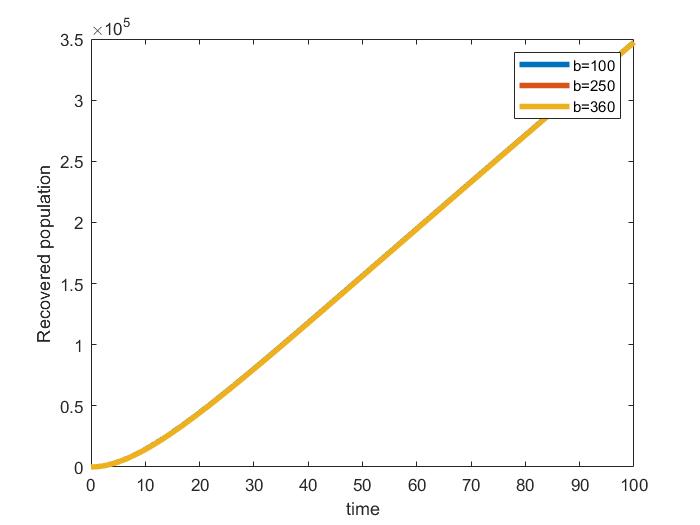}  
\end{subfigure}
\caption*{ }
\end{figure}

\begin{figure}[ht!]
\centering
\includegraphics[width=7cm]{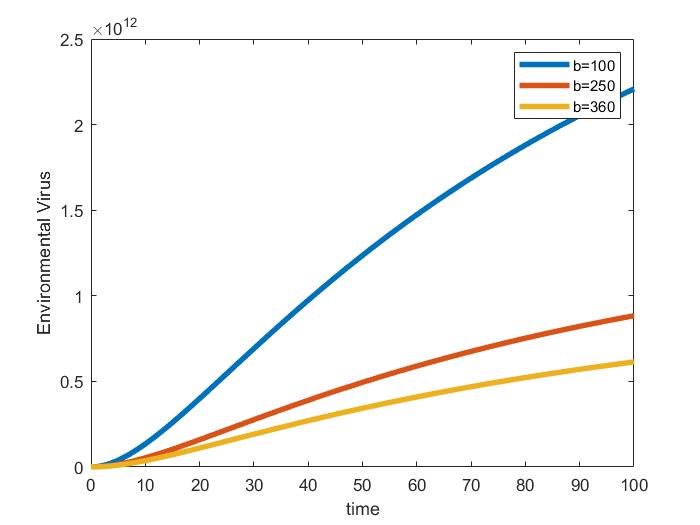}
\caption{Effects of variation of clearance rate of  virus  by immune system $(b)$.} \label{b}
\end{figure} \vspace{.35cm}

Figure \ref{b} depicts that as the rate of clearance of free virus particles increases, SARS-CoV-2 transmission in the community also decreases. Therefore, treatments that increase the rate of clearance of free virus particles in an infected individual have potential community-level benefits of reducing SARS-CoV-2 transmission at between-host scale apart from benefits to the infected individual.

\newpage

\begin{figure}
\begin{subfigure}
  \centering
  \includegraphics[width=7cm]{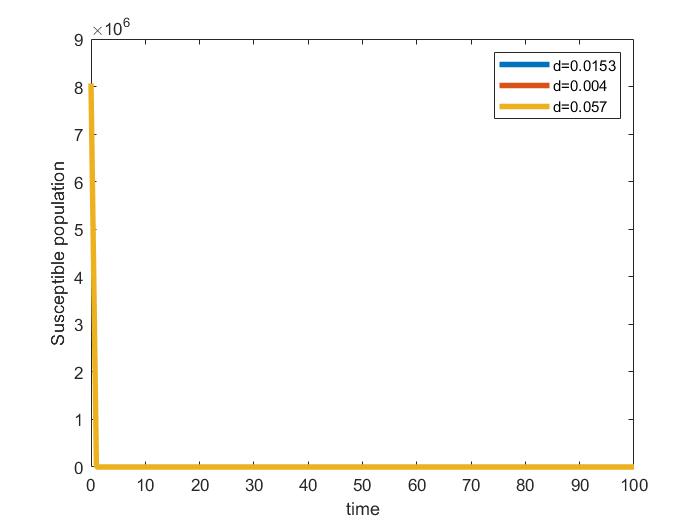}  
\end{subfigure}
\begin{subfigure}
  \centering
  \includegraphics[width=7cm]{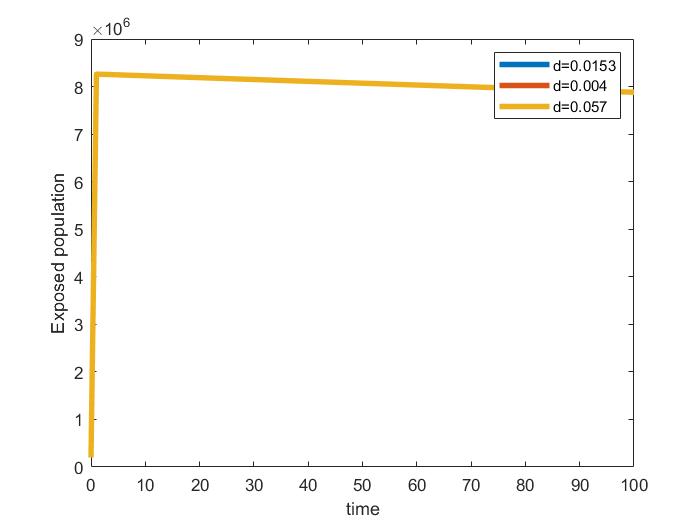}  
\end{subfigure}
\caption*{ }
\end{figure}

\begin{figure}
\begin{subfigure}
  \centering
  \includegraphics[width=7cm]{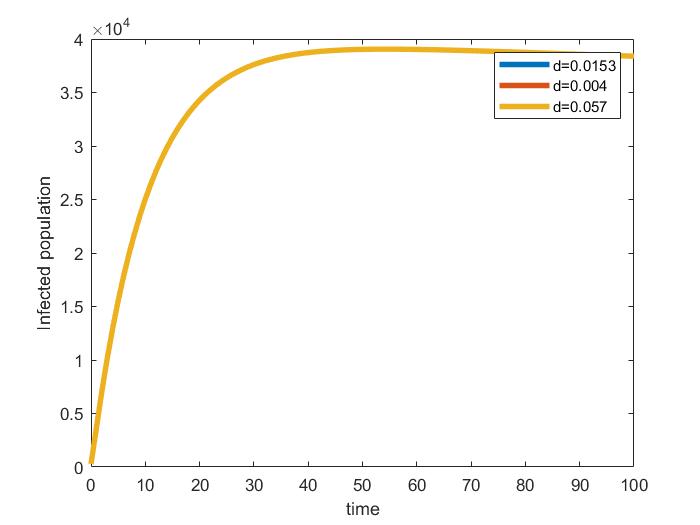}  
\end{subfigure}
\begin{subfigure}
  \centering
  \includegraphics[width=7cm]{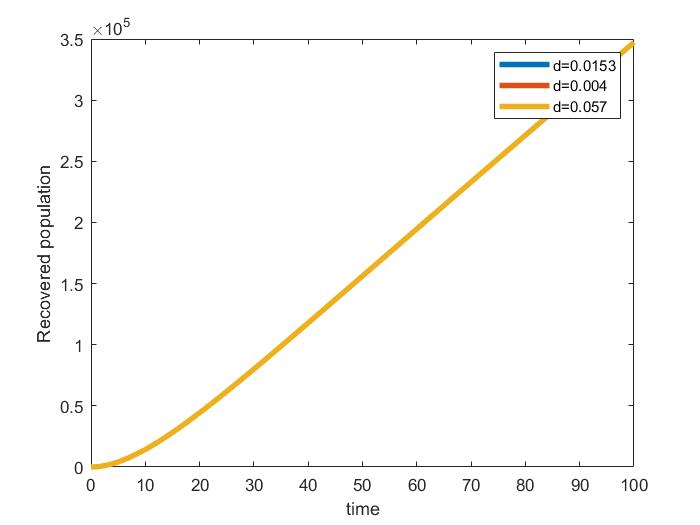}  
\end{subfigure}
\caption*{ }
\end{figure}

\begin{figure}[ht!]
\centering
\includegraphics[width=7cm]{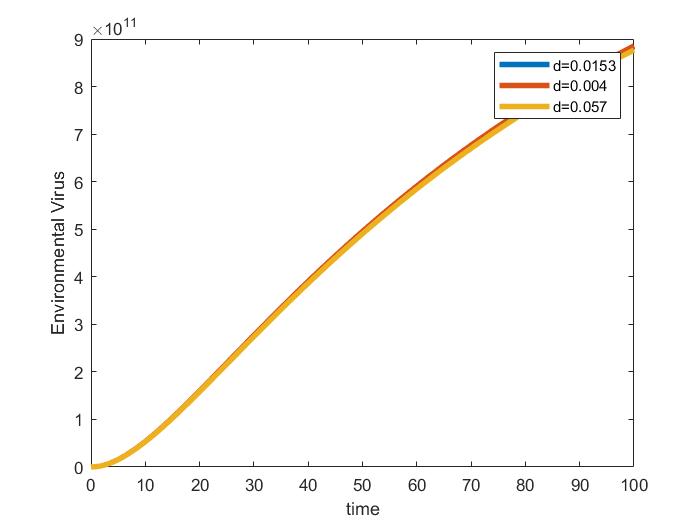}
\caption{Effects of variation of rate of killing of infected cells  by immune system $(d)$.} \label{d}
\end{figure} \vspace{.35cm}

Figure \ref{d} depicts that as the rate of killing of infected cells by immune system increases, SARS-CoV-2 transmission in the community also decreases slightly. Therefore, immunotherapies such as monoclonal antibodies that kill infected cells have potential community level benefits of reducing SARS-CoV-2 transmission at between-host scale apart from benefits to the infected individual.

\newpage

\section{Evaluating the comparative effectiveness of health interventions dealing with within-host and between-host scales} \vspace{.25cm}

In this section we do the comparative effectiveness studies.  We consider the following four health interventions dealing with within-host and between-host scales.

\begin{itemize}

\item[1.]  Antiviral drugs:

    a. 	Drugs that inhibit viral replication: Drugs such as  Arbidol, remdesivir, Lopinavir/Ritonavir
        inhibit viral replication in infected cells. So we choose $\alpha$ to be $\alpha(1-\epsilon).$

    b. 	Drugs that block virus binding to susceptible cells : Drugs such as Hydroxychloroquine (HCQ) does this job.
      So we choose $\beta$ to be $\beta(1-\gamma).$

\item[2.] Immunotherapies :  In this intervention   the
    rate of clearance of virus increases due to antibodies. This in turn  reduces the number of  infected cells. So we choose ($d_{1}$ + $d_{2}$ +$d_{3}$ + $d_{4}$ + $d_{5}$ + $d_{6}$) to be  ($d_{1}$ + $d_{2}$ +$d_{3}$ + $d_{4}$ + $d_{5}$ + $d_{6}$)($1+ \kappa$).
    
\item[3.] Environmental hygiene:   Decontamination from frequently touched spaces like door handles etc.,
    reduces environmental virus. So we choose $\pi$ to be $\pi(1 + \delta). $  
    
\item[4.] Generalized Social distancing: No mass gatherings, prayer meetings and educational institutions
    reduces contact with community viral load. So we choose $\eta_{w}$ to be $\eta_{w}(1 - \sigma).$ 
    
    \end{itemize}
    
$\mathcal{R}_{0}$ plays a crucial role in understanding the spread of infection in the individual and $\overline{V_{p}}$ determines the infectivity of virus in an individual. Taking these four health interventions into consideration, we now have modified basic reproductive number $ \mathcal{R}_{E}$ and modified virus count $\overline{V_{E}}$ of the endemic equilibrium to be

\begin{eqnarray*}
    \mathcal{R}_{E} &=& \frac{1}{2(\omega_{p}+\mu_{p})(\tau_{p}+\mu_{p})} \bigg[\eta_{p}\omega_{p}(1-\sigma) + \sqrt{(\eta_{p}\omega_{p})^{2}(1-\sigma)^2+\frac{4\eta_{w}\pi_{p}N_{e}\alpha_{h}\omega_{p}}{\mu_{p}\pi(1+\delta)}(\omega_{p}+\mu_{p})(\tau_{p}+\mu_{p})}\bigg] \\
    \overline{V_{E}} &=& \frac{N_{e}\alpha_{h}\pi_{p}}{\omega_{p}} \bigg[ \frac{\omega_{p}}{(\omega_{p}+\mu_{p})(\tau_{p}+\mu_{p})} - \frac{\pi(1+\delta)\mu_{p}}{\eta_{p}\mu_{p}\pi(1+\delta)(1-\sigma) + \eta_{w}N_{e}\alpha_{h}\pi_{p})}\bigg]  
\end{eqnarray*}

where, $N_{e}$ is the modified $N_{h}$ given by,

\begin{align*}
    N_{e} &= \frac{\mathfrak{R}_{e}}{\beta(1-\gamma)} [1-e^{-\mathfrak{R}_{e}}-\mathfrak{R}_{e} e^{-2\mathfrak{R}_{e}}], \\ 
    \mathfrak{R}_{e} &= \frac{\alpha\beta(1-\epsilon)(1-\gamma)S_{h}(s_{d_1})}{(b_1+b_2+b_3+b_4+b_5+b_6+\alpha_{h})(\mu+(d_1+d_2+d_3+d_4+d_5+d_6)(1+\kappa))}  
\end{align*}

We now do the comparative effectiveness study of these interventions by calculating the percentage reduction of $\mathcal{R}_{0}$ and $\overline{V_{p}}$  for single and multiple combination of these interventions at different efficacy levels such as (a) low efficacy of $0.3$, (b) medium efficacy of $0.6$, and (c)high efficacy of $0.9$. 

Percentage reduction of $\mathcal{R}_{0}$ and $\overline{V_{p}}$ are given by

Percentage reduction of $\mathcal{R}_{0}  = \bigg[ \frac{\mathcal{R}_{0} - \mathcal{R}_{E_j}}{\mathcal{R}_{0}} \bigg] \times 100$,

Percentage reduction of $\overline{V}_{p} = \bigg[ \frac{\overline{V_{p}} - \overline{V}_{E_j}}{\overline{V_{p}}} \bigg] \times 100$ 

where $j$ stands for $\epsilon,\delta,\kappa,\delta,\sigma$ or combinations thereof.

We now consider $20$ different combinations of these four health interventions. The comparative effectiveness is calculated and measured on a scale from $1$ to $20$ with $1$ denoting the lowest comparative effectiveness while $20$ denoting the highest comparative effectiveness.

\vspace{.25cm}

\begin{table}[htp!]
\begin{center}
\begin{tabular}{ | c | c | c | c | c | c | c | c |}
\hline
 \textbf{No.} & \textbf{Indicator} & \textbf{\%age} & \textbf{CEL} & \textbf{\%age} & \textbf{CEM} & \textbf{\%age} & \textbf{CEH} \\ 
 \hline \hline
 1 & $\mathcal{R}_{0}$ & 0  &  1  &  0  &  1  &  0  &  1 \\
 \hline
 2 & $\mathcal{R_{E_{\epsilon}}}$ & 16.34  &  12  &  36.75  &  12  &  68.38  &  12  \\
 \hline
 3 & $\mathcal{R_{E_{\gamma}}}$ & 0  &  2  &  0  &  2  &  0  &  2  \\ 
 \hline
 4 & $\mathcal{R_{E_{\kappa}}}$ & 1.79  &  4  &  3.47  &  4  &  5.09  &  4  \\ 
 \hline
 5 & $\mathcal{R_{E_{\delta}}}$ & 12.29  &  5  &  20.94  &  5  &  27.45  &  5  \\
 \hline
 6 & $\mathcal{R_{E_{\sigma}}}$ & 0.02  &  3  &  0.03  &  3  &  0.05  &  3  \\
 \hline
 7 & $\mathcal{R_{E_{\epsilon\delta\sigma}}}$ & 26.62  &  15  &  50  &  15  &  77.06  &  15  \\
 \hline
 8 & $\mathcal{R_{E_{\gamma\delta\sigma}}}$ & 12.29  &  7  &  20.94  &  7  &  27.45  &  7  \\ 
 \hline
 9 & $\mathcal{R_{E_{\kappa\delta\sigma}}}$ & 13.86  &  10  &  23.69  &  10  &  31.14  &  10  \\ 
 \hline
 10 & $\mathcal{R_{E_{\epsilon\delta}}}$ & 26.62  &  13  &  49.99  &  13  &  77.06  &  13  \\ 
 \hline
 11 & $\mathcal{R_{E_{\gamma\delta}}}$ & 12.29  &  6  &  20.94  &  6  &  27.45  &  6  \\ 
 \hline
 12 & $\mathcal{R_{E_{\kappa\delta}}}$ & 13.86  &  8  &  23.69  &  8  &  31.14  &  8  \\ 
 \hline
 13 & $\mathcal{R_{E_{\epsilon\gamma\delta}}}$ & 26.62  &  14  &  49.99  &  14  &  77.06  &  14  \\
 \hline
 14 & $\mathcal{R_{E_{\epsilon\kappa\delta}}}$ & 27.93  &  18  &  51.74  &  18  &  78.23  &  18  \\ 
 \hline
 15 & $\mathcal{R_{E_{\gamma\kappa\delta}}}$ & 13.86  &  9  &  23.69  &  9  &  31.14  &  9  \\
 \hline
 16 & $\mathcal{R_{E_{\epsilon\gamma\kappa\delta}}}$ & 27.93  &  17  &  51.74  &  17  &  78.23  &  17  \\ 
 \hline
 17 & $\mathcal{R_{E_{\epsilon\gamma\delta\sigma}}}$ & 26.62  &  16  &  50  &  16  &  77.06  &  16  \\ 
 \hline
 18 & $\mathcal{R_{E_{\epsilon\kappa\delta\sigma}}}$ & 27.93  &  20  &  51.74  &  20  &  78.23  &  20  \\ 
 \hline
 19 & $\mathcal{R_{E_{\gamma\kappa\delta\sigma}}}$ & 13.86  &  11  &  23.69  &  11  &  31.14  &  11  \\ 
 \hline
 20 & $\mathcal{R_{E_{\epsilon\gamma\kappa\delta\sigma}}}$ & 27.93  &  19  &  51.74  &  19  &  78.23  &  19  \\ 
 \hline\hline
\end{tabular}
\caption{Comparative effectiveness for $\mathcal{R_{0}}$}
\label{cet1}
\end{center}
\end{table}

\begin{table}[htp!]
\begin{center}
\begin{tabular}{ | c | c | c | c | c | c | c | c |}
\hline
 \textbf{No.} & \textbf{Indicator} & \textbf{\%age} & \textbf{CEL} & \textbf{\%age} & \textbf{CEM} & \textbf{\%age} & \textbf{CEH} \\ 
 \hline \hline
 1 & $\overline{V_{p}}$  & 0  &  1  &  0  &  1  &  0  &  1  \\
 \hline
 2 & $\overline{V_{E_{\epsilon}}}$ & 30.00 & 12 & 60.00 & 12 & 90.00 & 12  \\
 \hline
 3 & $\overline{V_{E_{\gamma}}}$ & 0  &  2  &  0  &  2  &  0  &  2  \\ 
 \hline
 4 & $\overline{V_{E_{\kappa}}}$ & 3.54  &  7  &  6.84  &  7  &  9.92  &  7  \\
 \hline
 5 & $\overline{V_{E_{\delta}}}$ & 0.01  &  5  &  0.01  &  5  &  0.01  &  5  \\
 \hline
 6 & $\overline{V_{E_{\sigma}}}$ & 0.01  &  3  &  0.01  &  3  &  0.01  &  3  \\ 
 \hline
 7 & $\overline{V_{E_{\epsilon\delta\sigma}}}$ & 30.00  &  15  &  60.00  &  15  &  90.00  &  15  \\
 \hline
 8 & $\overline{V_{E_{\gamma\delta\sigma}}}$ & 0.01  &  6  &  0.01  &  6  &  0.01  &  6  \\
 \hline
 9 & $\overline{V_{E_{\kappa\delta\sigma}}}$ & 3.54  &  11  &  6.84  &  11  &  9.92  &  11  \\
 \hline
 10 & $\overline{V_{E_{\epsilon\delta}}}$ & 30.00  &  13  &  60.00  &  13  &  90.00  &  13  \\ 
 \hline
 11 & $\overline{V_{E_{\gamma\delta}}}$ & 0.01  &  4  &  0.01  &  4  &  0.01  &  4  \\ 
 \hline
 12 & $\overline{V_{E_{\kappa\delta}}}$ & 3.54  &  9  &  6.84  &  9  &  9.92  &  9  \\ 
 \hline
 13 & $\overline{V_{E_{\epsilon\gamma\delta}}}$ & 30.00  &  14  &  60.00  &  14  &  90.00  &  14  \\  
 \hline
 14 & $\overline{V_{E_{\epsilon\kappa\delta}}}$ & 32.48  &  18  &  62.74  &  18  &  91,00  &  18  \\
 \hline
 15 & $\overline{V_{E_{\gamma\kappa\delta}}}$ & 3.54  &  8  &  6.84  &  8  &  9.92  &  8  \\
 \hline
 16 & $\overline{V_{E_{\epsilon\gamma\kappa\delta}}}$ & 32.48  &  17  &  62.74  &  17  &  90.99  &  17  \\ 
 \hline
 17 & $\overline{V_{E_{\epsilon\gamma\delta\sigma}}}$ & 30.00  &  16  &  60.00  &  16  &  90.00  &  16  \\ 
 \hline
 18 & $\overline{V_{E_{\epsilon\kappa\delta\sigma}}}$ & 32.48  &  20  &  62.74  &  20  &  90.99  &  20  \\
 \hline
 19 & $\overline{V_{E_{\gamma\kappa\delta\sigma}}}$ & 3.54  &  10  &  6.84  &  10  &  9.92  &  10  \\ 
 \hline
 20 & $\overline{V_{E_{\epsilon\gamma\kappa\delta\sigma}}}$ & 32.48  &  19  &  62.74  &  19  &  90.99  &  19  \\ 
 \hline\hline
\end{tabular}
\caption{Comparative effectiveness for $\overline{V_{p}}$}
\label{cet2}
\end{center}
\end{table}

The outcomes of the comparative effectiveness study suggest the following.
\begin{itemize}

\item[1.] When a single strategy is implemented, treatment with drugs such as Arbidol, remdesivir, Lopinavir/Ritonavir  that inhibits viral replication  show significant decrease of $\mathcal{R}_{0}$ relative to other four interventions at all efficacy levels.

\item[2.] Considering the severity of this pandemic, one single strategy is not sufficient to tackle this infection at the earliest. When environmental hygiene and generalized social distancing are implemented along with treatment of single drug, treating with drugs that inhibits viral replication performs better again at all efficacy levels.

\item[3.] Now that governments are accepting the fact that we have to live with the virus for long and planning for unlock strategies, generalized social distancing like closure of schools does not seem practical. By considering only environmental hygiene, along with the drugs that inhibits viral replication  seem to perform twice better than other drugs at all efficacy levels.

\item[4.] A combined strategy involving  treatment with drugs such as Arbidol, remdesivir, Lopinavir/Ritonavir that inhibits viral replication  and immunotherapies like monoclonal antibodies, along with environmental hygiene and generalized social distancing seems to perform the best among all combinations considered at all efficacy levels.

\end{itemize}

\section{Discussion and Conclusions} \vspace{.25cm}

 In this work  a novel nested multi scale model for COVID-19 is proposed and studied. We initially study the dynamics of this system and do the stability analysis. Later using the technique of  comparative effectiveness we study the efficacy of  four health interventions dealing with within-host and between-host scales. The results suggest that a combined strategy involving  treatment with drugs such as Arbidol, remdesivir, Lopinavir/Ritonavir that inhibits viral replication  and immunotherapies like monoclonal antibodies, along with environmental hygiene and generalized social distancing  proved to be the best and optimal in reducing the basic reproduction number and environmental virus at the population level.

 With a lot of research happening in the field of multi-drug therapy, our results offer some basic insights of their efficiency and effectiveness at population scale. These results can be helpful in public health measures and policies.
With more availability of data of within-host dynamics in COVID-19 patients, a better refined and comprehensive models can be framed based on this model that can more closer to real life situations. The multi scale modeling studies done here is the first of its kind for COVID-19.

\newpage

\bibliographystyle{amsplain}
\bibliography{bibliography}
\end{document}